\documentclass{crmp-l}

\newtheorem{theorem}{Theorem}[section]
\newtheorem{lemma}[theorem]{Lemma}

\theoremstyle{definition}
\newtheorem{definition}[theorem]{Definition}

\theoremstyle{remark}
\newtheorem{remark}[theorem]{Remark}

\numberwithin{equation}{section}

\newcommand{\mc}{\mathcal}

\newcommand{\tl}{\tilde}
\newcommand{\image}{\mbox{{\rm image}}}
\newcommand{\ovl}{\overline}
\newcommand{\supimage}{\mbox{\scriptsize{\rm image}}}
\newcommand{\sep}{\mbox{\scriptsize{\rm sep}}}
\begin{document}

\title[Separability and the BG normalization]
{Separability and the Birkhoff-Gustavson Normalization of
\\
the Perturbed Harmonic Oscillators with
\\
Homogeneous Polynomial Potentials}

%    Information for first author
\author{Yoshio Uwano}
%    Address of record for the research reported here
\address{Department of Applied Mathematics and Physics,
Kyoto University, Kyoto 606-8501, Japan}
%    Current address
%\curraddr{Department of Mathematics and Statistics,
%Case Western Reserve University, Cleveland, Ohio 43403}
\email{uwano@amp.i.kyoto-u.ac.jp}
%    \thanks will become a 1st page footnote.
\thanks{The author was supported in part by Grant-in-Aid
Scientific Research No.13660065, JSPS.}

%    General info
\date{January 1, 1994 and, in revised form, June 22, 1994.}

%\dedicatory{This paper is dedicated to our advisors.}


\begin{abstract}
In this paper, separability of the perturbed $2$-dimensional
isotropic harmonic oscillators with homogeneous polynomial potentials
is characterized from their Birkhoff-Gustavson (BG) normalization,
one of the conventional methods for {\it non-integrable} Hamiltonian systems.
\end{abstract}
\maketitle
%%%%%%%%%%%%%%%%%%%%%%%%%%%%%%%%%%%%%%%%%%%%%%%%%%%%%%%%%%%%%%%%%%%%%%%%
\section{Introduction}
The Bertrand-Darboux (BD) theorem is a very well-known theorem
established more than a century ago (\cite{Ber}, \cite{D}, \cite{MW}),
which characterizes separability and existence of
constants of motion quadratic in momenta of simple dynamical systems
on the Euclidean plane. As expected, the BD theorem has been playing
a key role of various studies on {\it integrable systems}
(see \cite{GPS}, \cite{H}, \cite{MW}, \cite{W} and references therein).
\par
On turning to {\it non-integrable systems}, the Birkhoff-Gustavson
(BG) normalization is known as one of the conventional
methods to them (\cite{M}):
For a given system feasible to be normalized,
the BG normalization provides a good account for the phase portrait
in the regular r{\' e}gime.
\par
Although directed to different characteristics of dynamical systems,
those well-known methods have encountered in the inverse problem of
the BG normalization which is posed by the author as follows
(\cite{UCRV}, \cite{U1}):
{\it For a given polynomial (or power series
\footnote{In discussing the BG normalization, we usually think of
power series in formal sense \cite{M}.}
) Hamiltonian in the BG normal form (BGNF),
identify all the possible Hamiltonians in polynomial
or in power series which share the given BGNF.}
In the inverse problem of the BG normalization of
the perturbed isotropic harmonic oscillators (PHOs) with
homogeneous polynomial
potentials of {\it degree-$3$}
\footnote{Throughout this paper, we deal with $2$-degrees of freedom
systems only.}
, the condition revealed in BD theorem (BDC) has come out as follows:
%%%
\begin{theorem}[\cite{U1}]
\label{3PHO-4PHO}
A PHO with homogeneous polynomial potential of degree-$3$
shares its BGNF up to degree-$4$ with a PHO of degree-$4$
if and only if the PHO of degree-$3$ satisfies the \lq generic' BDC
\footnote{See Eqs. (54a) and (55b) in \cite{U1}.}
.
The PHO of degree-$4$ corresponding to that PHO of degree-$3$
also satisfies the generic BDC.
\end{theorem}
%%%
From now on, the perturbed $2$-dimensional isotropic harmonic oscillator
with a homogeneous polynomial potential of degree-$\delta$ will be
abbreviated to as a \lq $\delta$-PHO'. The aim of this paper is to report
briefly that the following extension of Theorem~\ref{3PHO-4PHO} holds true
\footnote{More detailed discussion will be made in a pair of subsequent
papers, \cite{U2}.}
:
\begin{theorem}[main theorem]
\label{main}
For any odd $\delta$ greater than or equal to $3$, 
a $\delta$-PHO shares its BGNF up to degree-$(2\delta-2)$
with a $(2\delta-2)$-PHO if and only if the $\delta$-PHO is separable
within a rotation of Cartesian coordinates. The $(2\delta-2)$-PHO
sharing the BGNF with that $\delta$-PHO is also
separable within the same rotation of Cartesian coordinates.
\end{theorem}
%%%
The organization of this paper is outlined as follows.
In Section~2, the separability of the $\delta$-PHOs is studied
by applying the BD theorem to them. On associating the
$2 \times (\delta-1)$ matrix of the form
\begin{equation}
\label{M}
{\mc M}(K^{(\delta)})
=
\left(
\begin{array}{cccc}
v^{(\delta)}_0 - v^{(\delta)}_2 & v^{(\delta)}_1 - v^{(\delta)}_3 
& \cdots & v^{(\delta)}_{\delta-2} - v^{(\delta)}_{\delta}
\\
2v^{(\delta)}_1 & 2v^{(\delta)}_2 & \cdots & 2v^{(\delta)}_{\delta-1}
\end{array}
\right) 
\end{equation}
with the $\delta$-PHO Hamiltonian defined by
\begin{equation}
\label{Kd}
K^{(\delta)}(q,p)
=
\frac{1}{2}\sum_{j=1}^{2} (p_j^2 + q_j^2) + V^{(\delta)}(q)
\quad \mbox{and} \quad
V^{(\delta)}(q)=
\sum_{h=0}^{\delta}
 v^{(\delta)}_h \textstyle{{\delta \choose h}} q_1^{h} q_2^{\delta-h},
\end{equation}
the separability of the $\delta$-PHOs within rotations of Cartesian
coordinates is shown to be equivalent to
\begin{equation}
\label{deg-M}
\mbox{rank} \, {\mc M}(K^{(\delta)}) = 1.
\end{equation}
Since (\ref{deg-M}) with $\delta=3,4$ provides the \lq generic' BDC
for the $3$- and $4$-PHOs , the separability is taken as the extension
of the \lq generic' BDC.
In Section~3, the BG normalization of the $\delta$-PHOs is studied,
which provides a plausible reason to extend the relation between the
degrees, $3$ and $4$ of the PHOs in Theorem~\ref{3PHO-4PHO} to
$\delta$ and $2\delta-2$ of those in Theorem~\ref{main}.   
Section~4 is devoted to the proof of Theorem~\ref{main}:
The separability is shown to be sufficient for any $\delta$-PHO to
share its BGNF up to degree-$(2\delta-2)$ with a $(2\delta-2)$-PHO
in subsection~4.1, and is shown to be necessary in subsection~4.2.
%%%%%%%%%%%%%%%%%%%%%%%%
\section{The separability of the $\delta$-PHOs}
\subsection{The BD theorem for the $\delta$-PHOs}
To extend Theorem~\ref{3PHO-4PHO} to the PHOs of general degree,
we wish to understand more the meaning of the \lq generic' BDC
for the $3$- and $4$-PHOs${}^{3}$.
We start with applying the BD theorem to the
$\delta$-PHOs.
\begin{theorem}
\label{BD-theorem}
A $\delta$-PHO admits a first integral quadratic in momenta
if and only if it satisfies one of the followings:
\newline
{\rm (I)} \quad For odd $\delta \geq 5$, 
\begin{equation}
\label{BDC-odd}
\mbox{{\rm rank}} \, {\mc M}(K^{(\delta)}) = 1,
\end{equation}
where ${\mc M}(K^{(\delta)})$ is the matrix associated with the
$\delta$-PHO by (\ref{M}) and (\ref{Kd}).
\newline
{\rm (II)} \quad
For even $\delta \geq 4$, one of the following (\ref{BDC-even-a}) and
(\ref{BDC-even-b});
\begin{eqnarray}
\label{BDC-even-a}
&&
\mbox{{\rm rank}} \, {\mc M}(K^{(\delta)}) = 1,
\\
\label{BDC-even-b}
&&
\left\{
\begin{array}{ll}
v^{(\delta)}_{2h}= \left\{ \textstyle{{{\delta/2} \choose h}} / 
\textstyle{{\delta \choose 2h}} \right\} v^{(\delta)}_{0} \quad
\mbox{with} \quad v^{(\delta)}_{0} \ne 0
& \quad ( h =1, \cdots , \frac{\delta}{2})
\\ \noalign{\vskip 3pt}
v^{(\delta)}_{2h^{\prime}-1}=0 
& \quad ( h^{\prime} =1, \cdots , \frac{\delta}{2}).
\end{array}
\right.
\end{eqnarray}
\noindent
{\rm (III)} \quad For $\delta=3$, one of (\ref{BDC-3-a})
and (\ref{BDC-3-b});
%%
%%\numparts
%%
\begin{eqnarray}
\label{BDC-3-a}
&&
\mbox{{\rm rank}} \; {\mc M}(K^{(3)}) = 1,
\\
\label{BDC-3-b}
&&
\mbox{{\rm rank}}
\left(
\begin{array}{ccc}
7v_1 & -v_0 + 6v_2 & -2v_1+5v_3
\\
-5v_0+2v_2 & -6v_1+v_3 & -7v_2
\end{array}
\right)
=1.
\end{eqnarray}
\end{theorem}
%%%
\begin{proof}
The proof is made straightforward by writing down
explicitly the BDC (\cite{MW}, \cite{U1})
in terms of $v^{(\delta)}_{h}$s.
\par
The BDC: {\it There exist real-valued constants,
$
(\alpha, \beta, \beta^{\prime}, \gamma, \gamma^{\prime})
\ne
(0,0,0,0,0)
$,
for which the potential function $V(q)$ of a given natural dynamical
system satisfies}
%%%
\begin{eqnarray}
\label{BDC}
&&
\left(
\frac{\partial^2 V}{\partial q_2^2}
-\frac{\partial^2 V}{\partial q_1^2}
\right)
(-2\alpha q_1q_2-\beta^{\prime}q_2 -\beta q_1 + \gamma)
\\
\nonumber
&&
\phantom{xxxxxxx}
+2 \frac{\partial^2 V}{\partial q_1 \partial q_2}
(\alpha q_2^2 -\alpha q_1^2 +\beta q_2 -\beta^{\prime}q_1
+\gamma^{\prime})
\\
\nonumber
&&
\phantom{xxxxxxxxxxx}
+\frac{\partial V}{\partial q_1}
(6\alpha q_2 + 3\beta)
-
\frac{\partial V}{\partial q_2}
(6\alpha q_1 + 3\beta^{\prime})=0 .
%}
\end{eqnarray}
%%%%%%
\indent
(i) $\delta >3 $: \quad
On substituting 
\begin{equation}
\label{subst-pot}
V(q)=\frac{1}{2}\sum_{j=1}^{2}(p_j^2 +q_j^2) +V^{(\delta)}(q)
\end{equation}
with (\ref{Kd}) into (\ref{BDC}), the lhs, denoted by
${\mc L}^{(\delta)}$, of (\ref{BDC}) is calculated to be
\begin{equation}
\label{L>3}
{\mc L}^{(\delta)}
=
{\mc L}^{(\delta)}_{\delta} +{\mc L}^{(\delta)}_{\delta-1}
  +{\mc L}^{(\delta)}_{\delta-2} +{\mc L}^{(\delta)}_{1}
\end{equation}
with the homogeneous polynomial parts,
\begin{equation}
\label{L1>3}
{\mc L}^{(\delta)}_{1} = 3\beta q_1 - 3\beta^{\prime} q_2,
\end{equation}
\begin{eqnarray}
\label{Ld-2>3}
&& \phantom{=} 
{\mc L}^{(\delta)}_{\delta-2} = \delta(\delta-1) \sum_{h=0}^{\delta-2}  
\textstyle{{{\delta-2} \choose {h}}}
\{  (v^{(\delta)}_h -v^{(\delta)}_{h+2}) \gamma 
+ 2 v^{(\delta)}_{h+1} \gamma^{\prime} \}
q_1^h q_2^{\delta-2-h} 
\\
\nonumber
\phantom{{\mc L}^{(\delta)}_{\delta-2} }
&& =
\delta ( \delta-1) \times (\gamma, \, \gamma^{\prime}) {\mc M}(K^{(\delta)}) 
(q_2^{\delta-2}, \, \cdots \, , \,
 \textstyle{{{ \delta-2} \choose {h}}} \, q_1^h q_2^{\delta-2-h} , 
  \, \cdots , \, q_1^{\delta-2})^T ,
\end{eqnarray}
\begin{eqnarray}
\label{Ld-1>3}
&&
\phantom{=}
{\mc L}^{(\delta)}_{\delta-1} 
\\
\nonumber
&& = 
\beta \delta \Big[
  (2\delta+1)v_1^{(\delta)}q_2^{\delta-1}+
     \left\{ (\delta+2)v_{\delta}^{(\delta)} 
             - (\delta-1)v_{\delta-2}^{(\delta)} \right\} q_1^{\delta-1}  
\\
\nonumber
&& \phantom{=} 
+ \sum_{h=1}^{\delta-2}
  \left\{ \left( 3\delta \textstyle{{{\delta-1} \choose {h}}}
            -(\delta-1) \textstyle{{{\delta-2} \choose {h}}} \right)
             v^{(\delta)}_{h+1} 
-3(\delta-1) \textstyle{{{\delta-2} \choose {h-1}}} v^{(\delta)}_{h-1}
          \right\} q_1^h q_2^{\delta-1-h}
  \Big]
\\
\nonumber
&& \phantom{=} 
 - \beta^{\prime} \delta \Big[
   \left\{ (\delta+2)v_0^{(\delta)} - (\delta-1)v_{2}^{(\delta)} \right\}
            q_2^{\delta-1} +(2\delta+1)v_1^{(\delta)}q_1^{\delta-1} 
\\
\nonumber
&& \phantom{=} 
- \sum_{h=1}^{\delta-2}
\left\{
 \left( 3\delta \textstyle{{{\delta-1} \choose {h}}}
        -(\delta-1) \textstyle{{{\delta-2} \choose {h-1}}} \right)
        v^{(\delta)}_{h} 
 -3(\delta-1) \textstyle{{{\delta-2} \choose {h}}} v^{(\delta)}_{h+2}
        \right\} q_1^h q_2^{\delta-1-h} 
\Big] ,
\end{eqnarray}
%%%
and
%%%
\begin{eqnarray}
\label{Ld>3}
&&
{\mc L}^{(\delta)}_{\delta}
= 2\alpha \delta(\delta+2) 
\\
\nonumber
&& \phantom{{\mc L}^{(\delta)}_{\delta}=} 
\times
 \left[ 
v^{(\delta)}_1 q_2^{\delta} - v^{(\delta)}_{\delta-1} q_1^{\delta}
\displaystyle{
+ \sum_{h=1}^{\delta-1}
\left\{
\textstyle{{{\delta-1} \choose {h}}} v^{(\delta)}_{h+1}
-
\textstyle{{{\delta-1} \choose {h-1}}} v^{(\delta)}_{h-1}
\right\}
q_1^{h} q_2^{\delta-h}
}
\right] ,
\end{eqnarray}
%%%%%%
of degree-$1$, -$(\delta-2)$, -$(\delta-1)$ and -$\delta$, respectively
\footnote{The superscript ${}^T$ stands for the transpose throughout this
paper.}
. 
\par
(ii) $\delta=3$: \quad
Substituting (\ref{subst-pot}) with $\delta=3$ into (\ref{BDC}),
we obtain
\begin{equation}
\label{L=3}
{\mc L}^{(3)}
=
{\mc L}^{(3)}_{3} +{\mc L}^{(3)}_{2}+{\mc L}^{(3)}_{1}
\end{equation}
with
\begin{equation}
\label{L1=3}
{\mc L}^{(3)}_1
=
3\left\{  
(\gamma , \gamma^{\prime} )
{\mc M}(K^{(3)}) 
+
( -\beta^{\prime} , \beta )
\right\}
\left( \begin{array}{c} q_2 \\ q_1 \end{array} \right)
\end{equation}
and
\begin{equation}
\label{L2=3}
{\mc L}^{(3)}_2 
=
3(\beta, \beta^{\prime})
\left(
\begin{array}{ccc}
7v_1 & -v_0 + 6v_2 & -2v_1+5v_3
\\
-5v_0+2v_2 & -6v_1+v_3 & -7v_2
\end{array}
\right)
( q_2^2 , 2q_1q_2 , q_1^2)^T,
\end{equation}
where ${\mc L}^{(3)}_3$ is given by (\ref{Ld>3}) with $\delta=3$.
\par
From the explicit expression of ${\mc L}^{(\delta)}$ thus obtained,
we have Table~1, which classifies the possible choice of
$(\alpha, \beta , \beta^{\prime}, \gamma , \gamma^{\prime}) \ne 0$.

%%%%%
\begin{table}[h]
\label{tab-proof}
\caption{The BDC for the $\delta$-PHOs.} 
\begin{center}
%\item[]
\begin{tabular}{@{}lccc}
\noalign{\hrule height0.8pt}
& $\delta$:odd ($\geq 5$) & $\delta$:even ($\geq 4$) & $\delta=3$
\\
\hline
$\alpha=0$, $(\beta , \beta^{\prime})=(0,0)$,
$(\gamma , \gamma^{\prime}) \ne (0,0)$
& ($2.1$) & ($2.2$) & ($2.4$)
\\
$\alpha=0$, $(\beta , \beta^{\prime}) \ne (0,0)$,
$(\gamma , \gamma^{\prime}) \ne (0,0)$
& --- & --- & ($2.5$)
\\
$\alpha \ne 0$, $(\beta , \beta^{\prime}) = (0,0)$,
$(\gamma , \gamma^{\prime}) =(0,0)$
& --- & ($2.3$) & ---
\\
\noalign{\hrule height0.8pt}
\end{tabular}
\end{center}
\end{table}
%%%%%%%%%%
The derivation of Table~\ref{tab-proof} will be given in more detail
in \cite{U2}.
\end{proof}
%%%%%%%%%%%%%%%%%%%%%%%%%%%%
\subsection{The separability}
As expected from the BD theorem, the classification,
(\ref{BDC-odd})-(\ref{BDC-3-b}), of the BDC can be characterized
from the separability viewpoint.
\begin{theorem}
\label{BDC-sep}
A $\delta$-PHO is separable
\footnote{
As known well, a Hamiltonian system is said to be separable iff
the Hamilton-Jacobi equation for that system is separable.
}
within a rotation of Cartesian coordinates
if and only if (\ref{deg-M}) holds true.
\end{theorem}
%%%%
\begin{proof}
From the Hamilton-Jacobi equation
\begin{equation}
\label{HJ-Cart}
\frac{1}{2}\sum_{j=1}^{2}
\left(\frac{\partial S}{\partial q_j} \right)^2
+
V^{(\delta)}(q)
=E \qquad (E: \mbox{energy value})
\end{equation}
for the $\delta$-PHOs ($S(q)$: the generating function),
it is easy to see that the separation of (\ref{HJ-Cart})
with in rotations of Cartesian coordinates amounts to
that of $V^{(\delta)}(q)$. Accordingly, let us assume that
$V^{(\delta)}(q)$ is separated within a rotation
\begin{equation}
\label{kappa}
\kappa_{\psi} :
(q,p) \rightarrow ({\tl q} , {\tl p})=(\sigma (\psi) q , \sigma (\psi)p)
\end{equation}
with
\begin{equation}
\label{rot-mat}
\sigma (\psi) = \left(
\begin{array}{rr}
 \cos \psi & -\sin \psi \\ \sin \psi & \cos \psi
\end{array}
\right) \qquad (0 \leq \psi < 2\pi) .
\end{equation}
Namely, 
\begin{equation}
\label{tilde-V}
{\tl V}^{(\delta)}({\tl q})=V^{(\delta)}(\sigma({\psi})^{-1}{\tl q})
=
{\tl v}^{(\delta)}_0 {\tl q}_2^{\delta}
+
{\tl v}^{(\delta)}_{\delta} {\tl q}_1^{\delta},
\end{equation}
where $({\tl v}^{(\delta)}_0, {\tl v}^{(\delta)}_{\delta})\ne (0,0)$.
Equations (\ref{kappa})-(\ref{tilde-V}) are put together to yield
the relation,
%%%
\begin{equation}
\label{rel-v}
(4 \sin 2\psi) (v_h^{(\delta)}-v_{h+2}^{(\delta)})
- (\cos 2\psi)(2 v_{h+1}^{(\delta)})=0
\quad (h=0, \cdots, \delta-2) 
\end{equation}
among $v^{(\delta)}_{h}$s, which immediately implies (\ref{deg-M}).
The converse is easily shown by tracing back the discussion above.
\end{proof}
%%%%%%%%%%%
\begin{remark}
It is also possible to characterize the other classes of $\delta$-PHOs
subject to BDC listed in Theorem~\ref{BDC} from the separation of variables
viewpoint: The condition (\ref{BDC-even-b}) is shown to be equivalent to
the separability in the polar coordinates, and (\ref{BDC-3-b}) in a
off-centered parabolic coordinates (\cite{U2}). 
\end{remark}
%%%%%%%%%%%%
Since the \lq generic' BDC for the $3$-PHOs and the $4$-PHOs${}^{3}$
are equivalent to (\ref{deg-M}) with $\delta=3, 4$, we can thereby
look the separability within rotations upon as the \lq generic' BDC
for $\delta$-PHOs owing to Theorem~\ref{BDC-sep}. 
%%%%%%%%%%%%%%%%%%%%%%%%%%%%%%%%%%%%%%%%%%%%%%%%%%%%%%%%%%%%%%%%%%
\section{The BG normalization of the $\delta$-PHO}
In this section, we proceed the BG normalization of the $\delta$-PHOs,
which provides us with a key other than Theorem~\ref{BDC-sep} to
extend Theorem~\ref{3PHO-4PHO} to Theorem~\ref{main}.
\par
We start with describe the way how the $\delta$-PHO Hamiltonian
$K^{(\delta)}(q,p)$ is brought into the BGNF.
Let $G^{(\delta)}(\xi,\eta)$ be the BGNF of $K^{(\delta)}(q,p)$
and $W^{(\delta)}(q,\eta)$ be the generating function
\footnote{
$W^{(\delta)}(q,\eta)$ is said to be of the second-type since it
is a function of the \lq old' position variables $q$ and the \lq new'
momentum ones $\eta$ (\cite{G}).}
used for the BG normalization, both of which are expressed
in power-series form${}^{1}$
\begin{equation}
\label{Gd}
G^{(\delta)}(\xi , \eta ) = \frac{1}{2}\sum_{j=1}^{2} (\eta_j^2 + \xi_j^2)
+\sum_{k=3}^{\infty} G^{(\delta)}_k (\xi , \eta) 
\end{equation}
and
\begin{equation}
\label{Wd}
W^{(\delta)}(q, \eta)
=
\sum_{j=1}^{2}q_j \eta_j + \sum_{k=3}^{\infty} W^{(\delta)}_k (q, \eta),
\end{equation}
where $G^{(\delta)}_k(\xi,\eta)$ and $W^{(\delta)}_{k}(q,\eta)$ denote
the homogeneous polynomial parts of degree-$k$ of $G^{(\delta)}(\xi,\eta)$
and $W^{(\delta)}(q,\eta)$, respectively.
We normalize $K^{(\delta)}(q,p)$ by applying the canonical transformation,
\begin{equation}
\label{tau}
\tau : (q,p) \rightarrow (\xi,\eta) 
\quad
\mbox{{\rm with}}
\quad
p= \frac{\partial W^{(\delta)}}{\partial q}
\quad
\mbox{{\rm and}}
\quad
\xi = \frac{\partial W^{(\delta)}}{\partial \eta}
\end{equation}
associated with $W^{(\delta)}(q,\eta)$. Namely, $G^{(\delta)}(\xi,\eta)$
is determined by
\begin{equation}
\label{def-BG-G}
G^{(\delta)}
\left(  \frac{\partial W^{(\delta)}}{\partial \eta}, \eta \right)
=
K^{(\delta)}
\left( q , \frac{\partial W^{(\delta)}}{\partial q} \right) .
\end{equation}
%%%%%%%%%%%%%%%%%
\begin{definition}[The BGNF]
\label{G-BGNF}
Let $G^{(\delta)}(\xi,\eta)$ be the power-series${}^{1}$
(\ref{Gd}),
where each $G^{(\delta)}_k (\xi , \eta )$ is a homogeneous polynomial
of degree-$k$ ($k=3,4,\cdots$) in $(\xi , \eta)$
\footnote{The homogeneous part of degree-$2$ in $G^{(\delta)}(\xi,\eta)$
is always in the isotropic harmonic oscillator form, due to (\ref{Wd}).}
.
Then $G^{(\delta)}(\xi , \eta)$ is said to be in the BGNF up to degree-$\rho$
if and only if it satisfies
%%%%%%%%
\begin{equation}
\label{PD-vanish}
G^{(\delta)}_{k}(\xi,\eta) \in \ker D^{(k)}_{\xi,\eta}
\quad (k=3, \cdots , \rho),
\end{equation}
where $D^{(k)}_{\xi,\eta}$ is the restrict of the linear differential
operator
\begin{equation}
\label{D}
D_{\xi,\eta}
=
\sum_{j=1}^{2} \left( 
\xi_{j} \frac{\partial}{\partial \eta_{j}}
-
\eta_{j} \frac{\partial}{\partial \xi_{j}}
\right) 
\end{equation}
on the vector space of homogeneous polynomials of degree-$k$
in $(\xi,\eta)$.
\end{definition}
%%%%%%%%%%%%%%%%
\begin{remark}
The $D_{\xi,\eta}$ is understood as the Poisson derivation (\cite{A}),
$D_{\xi,\eta}
=\big\{ \frac{1}{2}\sum_{j=1}^{2} (\eta_j^2 + \xi_j^2) , \; \cdot \; \big\}
$,
associated with the isotropic harmonic oscillator.
\end{remark}
The ordinary problem of the BG normalization of the $\delta$-PHOs
is posed as follows
\footnote{For the inverse problem of the BG normalization,
see \cite{UCRV} and \cite{U2}.}
:
%%%%%%%%%%%%%%%
\begin{definition}[The ordinary problem of degree-$\rho$,
\cite{UCRV}, \cite{U1}]
\label{def-ord}
Bring a given $\delta$-PHO Hamiltonian $K^{(\delta)}(q,p)$ of the form
(\ref{Kd}) into the power series, $G^{(\delta)}(\xi,\eta)$, in the BGNF
up to degree-$\rho$ through (\ref{def-BG-G}), where generating function
$W^{(\delta)}(q,\eta)$ of the second-type in the form (\ref{Wd}) is chosen
to satisfy (\ref{def-BG-G}) and
%%%%%%%
\begin{equation}
\label{cond-W}
W^{(\delta)}_{k}(q,\eta) \in \image D^{(\delta)}_{q,\eta}
\quad (k=3,4,\cdots , \rho) .
\end{equation}
%%%%%%
The $D^{(k)}_{q,\eta}$ is defined by (\ref{D}) with $q$ in place of $\xi$.
%%%%
\end{definition}
%%%%%%%%%%%%%%%
\begin{remark}
\label{rem-cond-W}
The condition (\ref{cond-W}) is very crucial to ensure the uniqueness of
the outcome, $G^{(\delta)}(\xi, \eta)$, from $K^{(\delta)}(q,p)$
(\cite{UCRV}, \cite{U1}).
\end{remark}
%%%%%%
\par
We are now in a position to present an explicit expression of
the BGNF $G^{(\delta)}(\xi,\eta)$ of the $\delta$-PHO Hamiltonian.
A straightforward calculation of (\ref{def-BG-G}) shows the following:
\begin{lemma}
\label{BG-norm-2(d-1)}
The BGNF, $G^{(\delta)}(\xi,\eta)$, of the $\delta$-PHO Hamiltonian
$K^{(\delta)}(q,p)$ in the form (\ref{Kd}) takes the form
\begin{equation}
\label{Gd<=2(d-1)}
G^{(\delta)}(\xi,\eta)
=
\frac{1}{2}\sum_{j=1}^{2} (\eta_j^2 + \xi_j^2)
+ G^{(\delta)}_{\delta}(\xi,\eta) + G^{(\delta)}_{2\delta-2}(\xi,\eta)
+ o_{2\delta-1}(\xi , \eta),
\end{equation}
where $o_{2\delta-1}(\xi,\eta)$ denotes a power series in $(\xi,\eta)$
starting from degree-$(2\delta-1)$. 
The homogeneous polynomial part, $G^{(\delta)}_{\delta}(\xi,\eta)$,
of degree-$\delta$ is given by
%%%%%%%%%%%%%%%%
\begin{eqnarray}
\label{Gdd}
&&
\phantom{=}
G^{(\delta)}_{\delta} (\xi, \eta)=V^{(\delta)}{}^{\ker}(\xi, \eta)
\\
&&
=
\left\{
\begin{array}{ll}
0 &  \quad (\mbox{$\delta$: odd})
\\
\nonumber
\displaystyle{
2^{-\delta} \sum_{h=0}^{\delta}
}
 \textstyle{{\delta \choose h}}
 v^{(\delta)}_{h}
\displaystyle{
\left[
%%%%replaced on 1, May%%%%%%%%%%%%%%%%%
%%%%\sum_{m=0}^{h} \sum_{n=0}^{\delta-h}
%%%%%%%%%%%%%%%%%%%%%%%%%%%%%%%%%%%%%%%%
\sum_{m=M^{\flat}}^{M^{\sharp}} \sum_{n=N^{\flat}}^{N^{\sharp}}
\textstyle{{h \choose m}} \textstyle{{{\delta-h} \choose n}}
\zeta_1^{m} \zeta_2^{n}{\ovl \zeta}_1^{h-m} {\ovl \zeta}_2^{\delta-h-n}
\right]
}
& \quad (\mbox{$\delta$: even}),
\end{array}
\right.
\end{eqnarray}
%%%%%%%%%%%%%%%%
where $\zeta_j= \xi_j +i\eta_j  \quad (j=1,2)$, and
the ranges of the summation indices, $m$ and $n$, are determined by
%%%%%%%Added on 1, May%%%%%%%%%
\begin{equation}
\label{range-mn}
\left\{
\begin{array}{ll}
M^{\flat}=\max (0,h-\frac{\delta}{2}) 
& \quad 
M^{\sharp}=\min (h, \frac{\delta}{2}) 
\\
N^{\flat}=\max (0, \frac{\delta}{2}-h) 
& \quad 
N^{\sharp}=\min (\delta -h, \frac{\delta}{2}). 
\end{array}
\right.
\end{equation}
%%%%%%%%%%%%%%%%%%%%%%%%%%%%%%%%%%%%%
The superscript
$\, {}^{\ker}$ stands for taking the kernel component of $V^{(\delta)}(q)$
according to the action of $D^{(\delta)}_{q,\eta}$.
The homogeneous polynomial part, $G^{(\delta)}_{2\delta-2}(\xi,\eta)$,
of degree-$(2\delta-2)$ is calculated to be
%%%%%%%
\begin{equation}
\label{Gd-(2d-2)}
G^{(\delta)}_{2\delta-2}(\xi, \eta)
=
\sum_{m=0}^{2\delta-2}
\sum_{\ell =L^{\flat}}^{L^{\sharp}}
c^{(\delta)}_{m,\ell}
\zeta_1^{\ell}\zeta_2^{\delta-1-\ell}
{\ovl \zeta}_1^{m-\ell}{\ovl \zeta}_2^{\delta-1-m+\ell}
\end{equation}
with
\begin{eqnarray}
\label{coeff-Gd}
c^{(\delta)}_{m, \ell}
&=&
\frac{2\delta^2}{4^\delta}
\sum_{j=J^{\flat}}^{J^{\sharp}}
\left[
\textstyle{
{ {\delta-1} \choose j}
{ {\delta-1} \choose {m-j}}
}
(v^{(\delta)}_j v^{(\delta)}_{m-j} 
+ v^{(\delta)}_{j+1}v^{(\delta)}_{m-j+1})
\right.
\\ \noalign{\vskip 6pt}
\nonumber
&& 
\left.
\displaystyle{
\times
\sum_{k=K^{\flat}}^{K^{\sharp}} \sum_{h=H^{\flat}}^{H^{\sharp}}
\frac{
{ j \choose k }
{ {\delta-1-j} \choose h }
{ {m-j} \choose {\ell -k}}
{ {\delta-1-(m-j)} \choose {\delta-1-\ell-h}}
}
{\{(2(k+h+1)-\delta \} \{2(k+h)-\delta \}}
}
\right] .
\end{eqnarray}
The ranges of the indices, $h$, $j$, $k$ and $\ell$,
in (\ref{Gd-(2d-2)}) and (\ref{coeff-Gd}) are determined by
\begin{equation}
\label{sum-ind}
\left\{
\begin{array}{ll}
H^{\flat}= \max (0, (m-j)-\ell) & \quad
H^{\sharp} = \min (\delta-1-\ell , \delta-1-j) 
\\
J^{\flat}= \max (0, m-(\delta-1)) & \quad
J^{\sharp} = \min (m , \delta-1)
\\
K^{\flat}= \max (0, \ell-(m-j)) & \quad
K^{\sharp} = \min (j , \ell)
\\
L^{\flat}= \max (0, m-(\delta-1)), & \quad
L^{\sharp} = \min (m , \delta-1)
\end{array}
\right.
\end{equation}
%%%%
respectively. 
%%%%%%%%%%%%%
\end{lemma}
%%%%
The proof is outlined in Appendix~\ref{proof-lemma}
and will be given in more detail in \cite{U2}.
From Lemma~\ref{BG-norm-2(d-1)}, we obtain the following:
%%%%%%%%%%%%%%%%%%%%%%%%
\begin{theorem}
\label{Gd1=Gd2}
Let $\delta_1$ and $\delta_2$ be any integers subject to
\begin{equation}
\label{assume-delta}
3 \leq \delta_1 < \delta_2.
\end{equation}
If the BGNF, $G^{(\delta_1)}(\xi,\eta)$,
of a $\delta_1$-PHO Hamiltonian $K^{\delta_1}(q,p)$ coincides with
the BGNF, $G^{(\delta_2)}(\xi,\eta)$, of a $\delta_2$-PHO Hamiltonian
$K^{(\delta_2)}(q,p)$ up to degree-$\delta_2$, then $\delta_1$ and $\delta_2$
have to satisfy
\begin{equation}
\label{cond-delta}
\mbox{$\delta_1$ : an odd integer}
\quad
\mbox{and}
\quad
\delta_2=2\delta_1-2.
\end{equation}
\end{theorem}
%%%%%%%%%%%%%%%%%%%%%%
\begin{proof}
Since the coincidence of the BGNFs is expressed as
\begin{equation}
\label{alt-coincide}
G^{(\delta_1)}(\xi,\eta)-G^{(\delta_2)}(\xi,\eta)=o_{\delta_2+1}(\xi, \eta),
\end{equation}
we obtain
\begin{equation}
\label{coincide-d1}
G^{(\delta_1)}_{\delta_1}(\xi,\eta)=G^{(\delta_2)}_{\delta_1}(\xi,\eta)=0
\end{equation}
from Lemma~\ref{BG-norm-2(d-1)} as a necessary condition for
(\ref{alt-coincide}). Equation (\ref{coincide-d1}) is put together with
(\ref{Gdd}) to yield the first condition in (\ref{cond-delta}).
We derive the second one in turn.
On recalling Lemma~\ref{BG-norm-2(d-1)} again,
the lowest non-vanishing homogeneous
polynomial part of $G^{(\delta_1)}(\xi,\eta)$ turns out to be
$G^{(\delta_1)}_{2\delta_1-2}(\xi,\eta)$ if $\delta_1$ is odd,
while $G^{(\delta_2)}_{\delta_2}(\xi,\eta)$ is the lowest one of
$G^{(\delta_2)}(\xi,\eta)$. This shows the second equation of
(\ref{cond-delta}).
\end{proof}
Now that we have a pair of key Theorems~\ref{BDC-sep} and \ref{Gd1=Gd2},
we are led to pose Theorem~\ref{main} as an extension of
Theorem~\ref{3PHO-4PHO}.
%%%%%%%%%%%%%%%%%%%%%%%%%%%%%%%%%%%%%%%%%%%%%%%%%%%%%%%%%
\section{Proof of the main theorem}
In this section, the proof of Theorem~\ref{main}
is outlined. Throughout this section, we assume
$\delta$ ($\geq 3$) to be odd.
%%%%%%%
\subsection{Part I: the separability as a sufficiency}
This subsection is devoted to show that the separability is sufficient
for a $\delta$-PHO to share its BGNF with a $(2\delta-2)$-PHO
up to degree-$(2\delta-2)$. 
\begin{remark}
\label{two-meanings}
Recalling the proof of Theorem~\ref{BDC-sep}, we see that the separability
of the $\delta$-PHOs${}^5$ within rotations of Cartesian
coordinates is equivalent to the separability of their Hamiltonians. 
We will use those equivalent expressions properly according to circumstances
henceforce.
\end{remark}
%%%%%%%
\subsubsection{The BG normalization of the $\delta$-PHOs in separate form}
Let the $\delta$-PHO be associated with the Hamiltonian in separate form,
\begin{equation}
\label{sep-Kd}
K^{(\delta)}_{\sep}(q,p)
=
\frac{1}{2}\sum_{j=1}^{2} (p_j^2 + q_j^2 )
+
(v^{(\delta)}_{\sep , 0}q_2^{\delta} 
+ v^{(\delta)}_{\sep , \delta}q_1^{\delta}).
\end{equation}
Then, applying (\ref{Gd-(2d-2)}) and (\ref{coeff-Gd}) to
$K^{(\delta)}_{\sep}(q,p)$, we obtain the BGNF of
$K^{(\delta)}_{\sep}(q,p)$, denoted by $G^{(\delta)}_{\sep}(\xi, \eta)$,
to be in the following separate form,
\begin{eqnarray}
\label{sep-Gd}
&&
\displaystyle{
G^{(\delta)}_{\sep}(\xi, \eta)
=\frac{1}{2}
\sum_{j=1}^{2}(\eta_j^2 + \xi_j^2)
+\left\{
\frac{2\delta^2}{4^{\delta}}\sum_{n=0}^{\delta-1}
\frac{{{\delta-1} \choose n}^{2}}{\{2(n+1)-\delta\}(2n-\delta)} 
\right\}
}
\\
\nonumber
&&
\phantom{G^{(\delta)}_{\sep}(\xi, \eta)==}
\times
\{v^{(\delta)}_{\sep , 0}{}^2 (\zeta_2{\ovl \zeta_2})^{\delta-1}
  + v^{(\delta)}_{\sep , \delta}{}^2 (\zeta_1{\ovl \zeta_1})^{\delta-1} \}
+ o_{2\delta-1}(\xi,\eta),
\end{eqnarray}
where $\zeta_j=\xi +\eta_j$ ($j=1,2$).
Recalling (\ref{Gd<=2(d-1)}) and (\ref{Gdd}) with $2\delta-2$
in place of $\delta$, we can find the unique
$(2\delta-2)$-PHO Hamiltonian
\begin{eqnarray}
\label{sep-K-(2d-2)}
&&
\displaystyle{
K^{(2\delta-2)}_{\sep}(q,p)
=
\frac{1}{2}\sum_{j=1}^{2} (p_j^2 + q_j^2 )
+\left\{ \frac{\delta^2}{2 {{2\delta-2} \choose {\delta-1}}}
\sum_{n=0}^{\delta-1}
\frac{{{\delta-1} \choose n}^{2}}{\{2(n+1)-\delta\}(2n-\delta)} 
\right\}
}
\\
\nonumber
&&
\phantom{K^{(2\delta-2)}_{\sep}(q,p)==}
\times
\{ (v^{(\delta)}_{\sep , 0})^2 q_2^{2\delta-2}
  + (v^{(\delta)}_{\sep , \delta})^2 q_1^{2\delta-2} \}
\end{eqnarray}
%%%
in separate form,
whose BGNF coincides with $G^{(\delta)}_{\sep}(\xi, \eta)$
up to degree-$(2\delta-2)$. To summarize, we have the following.
%%%
\begin{lemma}
\label{sep-dPHO-(2d-2)PHO}
For any $\delta$-PHO Hamiltonian $K^{(\delta)}_{\sep}(q,p)$
in separate form (\ref{sep-Kd}),
there exists the unique $(2\delta-2)$-PHO Hamiltonian
$K^{(2\delta-2)}_{\sep}(q,p)$ in separate form (\ref{sep-K-(2d-2)})
which shares the BGNF $G^{(\delta)}_{\sep}(\xi,\eta)$
up to degree-$(2\delta-2)$ with $K^{(\delta)}_{\sep}(q,p)$.
\end{lemma}
%%%%%%%%%%%%%%%%%%%%%
\subsubsection{Proof of the sufficiency}
A key to prove the sufficiency is the commutativity of the BG normalization
and the rotations of Cartesian coordinates:
%%%
\begin{lemma}
\label{commute}
Let $G^{(\delta)}(\xi, \eta)$ be the BGNF of a $\delta$-PHO Hamiltonian
$K^{(\delta)}(q,p)$ up to degree-$(2\delta-2)$,
which associates with the generating function
$W^{(\delta)}(q,\eta)$ (see (\ref{def-BG-G})).
Let ${\tl G}^{(\delta)}({\tl \xi}, {\tl \eta})$,
${\tl K}^{(\delta)}({\tl q}, {\tl p})$
and ${\tl W}^{(\delta)}({\tl q}, {\tl \eta})$ be the power series
defined by
\begin{eqnarray}
\nonumber
&&
{\tl K}^{(\delta)}({\tl q}, {\tl p})
=
K^{(\delta)}(\sigma^{-1}(\psi){\tl q}, \sigma^{-1}(\psi){\tl p})
\\
\label{rot-KGW}
&&
{\tl G}^{(\delta)}({\tl \xi}, {\tl \eta})
=
G^{(\delta)}(\sigma^{-1}(\psi){\tl \xi}, \sigma^{-1}(\psi){\tl \eta})
\\
\nonumber
&&
{\tl W}^{(\delta)}({\tl q}, {\tl \eta})
=
W^{(\delta)}(\sigma^{-1}(\psi){\tl q}, \sigma^{-1}(\psi){\tl \eta})  ,
\end{eqnarray}
where $\sigma(\psi)$ is defined by (\ref{rot-mat}).
Then ${\tl G}^{(\delta)}({\tl \xi}, {\tl \eta})$ is the BGNF of
the $\delta$-PHO Hamiltonian ${\tl K}^{(\delta)}({\tl q}, {\tl p})$
up to degree-$(2\delta-2)$,
which is brought through the canonical transformation,
$({\tl q}, {\tl p}) \rightarrow ({\tl \xi}, {\tl \eta})$,
generated by ${\tl W}^{(\delta)}({\tl q}, {\tl \eta})$.
\end{lemma}
\begin{proof}
Due to the orthogonality of $\sigma(\psi)$, it is easily confirmed from
(\ref{rot-KGW}) that ${\tl K}^{(\delta)}({\tl q}, {\tl p})$, 
${\tl G}^{(\delta)}({\tl \xi}, {\tl \eta})$ and
${\tl W}^{(\delta)}({\tl q}, {\tl \eta})$ are other
$\delta$-PHO Hamiltonian, BGNF up to degree-$(2\delta-2)$
and generating function of the second-type
(cf. (\ref{Kd}), (\ref{Gd}) and (\ref{Wd})), respectively.
Further, the orthogonality of $\sigma(\psi)$ is put together with
(\ref{rot-KGW}) to yield the equation,
\begin{equation}
\label{def-tl-G}
{\tl K}^{(\delta)}
({\tl q}, \frac{\partial {\tl W}^{(\delta)}}{\partial {\tl q}})
=
{\tl G}^{(\delta)}
(\frac{\partial {\tl W}^{(\delta)}}{\partial {\tl \eta}}, {\tl \eta}),
\end{equation}
from (\ref{def-BG-G}), so that ${\tl G}({\tl \xi}, {\tl \eta})$ is
the BGNF of ${\tl K}({\tl q}, {\tl p})$. This completes the proof.
\end{proof}
%%%%%%%%%%%%%%%%%%%%%%
We are at the final stage to prove the sufficiency of the separability
in Theorem~\ref{main} now.
Let us assume that $K^{(\delta)}(q,p)$ is separable within a rotation
of Cartesian coordinates: Namely, there exists the transformation
$\kappa_{\psi}$ with a suitable $\psi \in [0,2\pi)$ (see (\ref{kappa})
and (\ref{rot-mat})) which brings
$K^{(\delta)}(q,p)$ to $K^{(\delta)}_{\sep}({\tl q}, {\tl p})$
through
\begin{equation}
\label{Kd->sep-Kd}
K^{(\delta)}_{\sep}({\tl q},{\tl p})
=
K^{(\delta)}(\sigma (\psi)^{-1}{\tl q}, \sigma (\psi)^{-1}{\tl p}),
\end{equation}
where $K^{(\delta)}_{\sep}({\tl q},{\tl p})$ takes the separate form
(\ref{sep-Kd}) with $({\tl q},{\tl p})$ in place of $(q,p)$.
Then on applying Lemma~\ref{commute} to the pair,
$K^{(\delta)}_{\sep}({\tl q}, {\tl p})$ and
$G^{(\delta)}_{\sep}({\tl \xi}, {\tl \eta})$,
the BGNF $G^{(\delta)}(\xi, \eta)$ of $K^{(\delta)}(q,p)$ up
to degree-$(2\delta-2)$ is given by
\begin{equation}
\label{sep-Gd->Gd}
G^{(\delta)}(\xi, \eta)
=
G^{(\delta)}_{\sep}(\sigma(\psi)\xi, \sigma (\psi) \eta).
\end{equation}
Further, according to Lemma~\ref{sep-dPHO-(2d-2)PHO}, we can find uniquely
the $(2\delta-2)$-PHO Hamiltonian in separate form, say
$K^{(2\delta-2)}_{\sep}({\tl q}, {\tl p})$, sharing the BGNF
$G^{(\delta)}_{\sep}({\tl \xi},{\tl \eta})$ up to degree-$(2\delta-2)$
with $K^{(\delta)}_{\sep}({\tl q},{\tl p})$.
On defining the $(2\delta-2)$-PHO Hamiltonian
$K^{(2\delta-2)}(q,p)$ and the BGNF $G^{(2\delta-2)}(\xi,\eta)$ by
\begin{equation}
\label{K-G-(2d-2)-other}
\begin{array}{l}
\displaystyle{
K^{(2\delta-2)}(q,p)=K^{(2\delta-2)}_{\sep}(\sigma(\psi)q, \sigma(\psi)p)
}
\\
\displaystyle{
G^{(2\delta-2)}(\xi,\eta)=
G^{(\delta)}_{\sep}(\sigma(\psi)\xi, \sigma(\psi)\eta),
}
\end{array}
\end{equation}
Lemma~\ref{commute} shows that $G^{(2\delta-2)}(\xi,\eta)$ is the BGNF
of $K^{(2\delta-2)}(q,p)$ up to degree-$(2\delta-2)$.
Equations (\ref{sep-Gd->Gd}) and (\ref{K-G-(2d-2)-other})
are put together to show the coincidence of $G^{(2\delta-2)}(\xi,\eta)$
with $G^{(\delta)}(\xi,\eta)$ up to degree-$(2\delta-2)$.
To summarize, we have the following.
\begin{theorem}
\label{sufficient}
Let $\delta$ be an odd integer greater than or equal to $3$.
If a $\delta$-PHO Hamiltonian is separable (see Remark~\ref{two-meanings})
within a rotation of Cartesian coordinates, there exists the unique
$(2\delta-2)$-PHO Hamiltonian which shares the same BGNF up
to degree-$(2\delta-2)$ with the $\delta$-PHO. The $(2\delta-2)$-PHO
is also separable within the same rotation of Cartesian coordinates.
\end{theorem}
%%
%%%%%%%%%%%%%%%%%%%%%%%%%%%%%%%%%%%%%%%%%%%%%%%%%
\subsection{Part~II: the separability as a necessity}
This subsection is devoted to prove that the sfeparability is a necessary
condition. Let us recall Lemma~\ref{BG-norm-2(d-1)} and equate
$G^{(\delta)}(\xi,\eta)$ with $G^{(2\delta-2)}(\xi,\eta)$ up to
degree-$(2\delta-2)$. 
As a necessary and sufficient condition for the equation thus obtained,
we have a number of equations
\begin{equation}
\label{nsc}
c^{(\delta)}_{m, \ell}=c^{(2\delta-2))}_{m, \ell}
\qquad (0 \leq m \leq 2\delta-2, \; L^{\flat} \leq \ell \leq L^{\sharp}).
\end{equation}
Since the number of equations in (\ref{nsc}) is so many and
since we have already shown the separability as a sufficiency,
it would not be so smart to study (\ref{nsc}) for all the
pairs of subscripts, $(m, \ell)$s. Hence as necessary condition
for (\ref{nsc}), we consider
\begin{equation}
\label{equate-coeff}
c^{(\delta)}_{m, 1} {\tl c}^{(2\delta-2)}_{m, 0}
=
c^{(\delta)}_{m, 0} {\tl c}^{(2\delta-2)}_{m, 1}
\qquad
(m=2,3, \cdots , \delta-1)
\end{equation}
with
\begin{equation}
\label{tilde-c(2(d-1))}
{\tl c}^{(2\delta-2)}_{m, \ell}
=
4^{1-\delta} 
\textstyle{
{ {2\delta-2)} \choose m}
{ m \choose \ell }
{ {2\delta-2-m} \choose {\delta-1-\ell}}
} \qquad (\ell=0,1).
\end{equation}
Note that we have 
$c^{(2\delta-2)}_{m, \ell} = v^{(2\delta-2)}_m {\tl c}^{(2\delta-2)}_{m, \ell}$
for $\ell =0,1$.
\par
We wish to draw (\ref{deg-M}) as a necessary condition from
(\ref{equate-coeff}).
To do this, it is useful to prepare the notation,
\begin{eqnarray}
%\fl
\label{U}
&&
U^{(\delta,m)}_{j}
=
(v^{(\delta)}_j v^{(\delta)}_{m-j} + v^{(\delta)}_{j+1} v^{(\delta)}_{m-j+1})
\\
\nonumber
&&
\phantom{U^{(\delta,m)}_{j}=}
-
(v^{(\delta)}_{j+1} v^{(\delta)}_{m-(j+1)} + v^{(\delta)}_{(j+1)+1} v^{(\delta)}_{m-(j+1)+1})
\quad
(j=0, 1, \dots , m-1).
\end{eqnarray}
Using (\ref{Gdd}), (\ref{Gd-(2d-2)}) and (\ref{U}),
we can put (\ref{equate-coeff}) into the form
\begin{eqnarray}
%\fl
\label{B-U}
&&
\left( \sum_{n=0}^{m} B_{n}^{(\delta,m)} \right)
(v^{(\delta)}_{m} v^{(\delta)}_{0}
 + v^{(\delta)}_{m+1} v^{(\delta)}_{1})
\\
\nonumber
&&
\phantom{xxxxxxxx}
+ \sum_{j=0}^{m-1}
\left( \sum_{n=0}^{j} B_{n}^{(\delta,m)} \right)
U^{(\delta,m)}_{j} =0
\qquad
(m=2, \cdots , \delta-1),
\end{eqnarray}
where $B_{n}^{(\delta,m)}$s are defined to be
%%%%%%%%%%%%%%%%%%%%%%%%%
\begin{eqnarray}
%\fl
\label{B}
&&
\left( (\delta-1)^2 
\textstyle{{\delta-2} \choose {m-1}} 
\right)^{-1} B_{n}^{(\delta,m)}
\\ \noalign{\vskip 9pt}
\nonumber
&&
\qquad = \left\{
\begin{array}{ll}
\displaystyle{
\sum_{k=m-n-1}^{\delta-n-1} 
\frac{
 {{m-1} \choose n} {{\delta-m} \choose {n+k+1-m}} {{\delta-2} \choose {k}}
}
{ \{ 2(k+1)-\delta \} (2k-\delta) }
}
\\ \noalign{\vskip 3pt}
\displaystyle{
\phantom{+++} + \sum_{k=m-n-1}^{\delta-n-1} 
\frac{
 {{m-1} \choose {n-1}} {{\delta-m} \choose {n+k+1-m}} {{\delta-2} \choose {k}}
}
{ \{ 2(k+2)-\delta \} \{2(k+1)-\delta \} }
}
\\ \noalign{\vskip 3pt}
\displaystyle{
\phantom{++++++} - \sum_{k=m-n}^{\delta-h-1} 
\frac
{
  {{m} \choose {n}} 
  {{\delta-m-1} \choose {n+k-m}} 
  {{\delta-1} \choose {k}}
}
{
  \{ 2(k+1)-\delta \} (2k-\delta)
}
} 
\quad (n \ne 0,m),
\\
\noalign{\vskip 12pt}
\displaystyle{
\sum_{k=m-1}^{\delta-2} 
\frac{
{{\delta-m} \choose {k+1-m}} {{\delta-2} \choose {k}}
}
{ \{ 2(k+1)-\delta \} (2k-\delta) }
} 
\\ \noalign{\vskip 3pt}
\displaystyle{
\phantom{---} - \sum_{k=m}^{\delta-1} 
\frac{
{{\delta-m-1} \choose {k-m}} {{\delta-1} \choose {k}}
}
{ \{ 2(k+1)-\delta \} (2k-\delta) }
} 
\quad (n =0),
\\
\noalign{\vskip 12pt}
\displaystyle{
\sum_{k=0}^{\delta-m-1} 
\frac{
{{\delta-m} \choose {k+1}} {{\delta-2} \choose {k}}
}
{ \{ 2(k+2)-\delta \} \{ 2(k+1)-\delta \} }
} 
\\ \noalign{\vskip 3pt}
\displaystyle{
\phantom{---} - \sum_{k=0}^{\delta-m-1} 
\frac{
{{\delta-m-1} \choose {k+1}} {{\delta-1} \choose {k}}
}
{ \{ 2(k+1)-\delta \} (2k-\delta) }
} 
\quad (n =m) .
\end{array}
\right.
%\end{array}
\end{eqnarray}
The definition (\ref{B}) of $B^{(\delta,m)}_{n}$s and several well-known
formula for the binomial coefficients are put together to show
\begin{equation}
\label{vanish-sum-B}
\sum_{n=0}^{m} B_{n}^{(\delta,m)} =0
\quad (m=2, \cdots , \delta-1)
\end{equation}
%%%
(see Appendix~\ref{proof-vanish}), so that we can put (\ref{B-U}) in the form
\begin{equation}
\label{B-U-cont}
\sum_{j=0}^{m-1}
\left( \sum_{n=0}^{j} B_{n}^{(\delta,m)} \right)
U^{(\delta,m)}_{j} =0
\quad
(m=2, \cdots , \delta-1).
\end{equation}
%%%%
We verify (\ref{B-U-cont}) further by
characterizing $U^{(\delta,m)}_{j}$
as the minors of ${\mc M}(K^{(\delta)})$ as follows.
Denoting by $\Delta^{(\delta)}_{a b}$ the minor of
${\mc M}(K^{(\delta)})$ consisting of its $a$- and $b$-th columns
($1 \leq a < b \leq \delta-1$), we have
\begin{equation}
\label{U-Delta}
U^{(\delta,m)}_{j}
=
\left\{
\begin{array}{ll}
\Delta^{(\delta)}_{j+1 \; m-j} & \quad (j=0, \cdots , [m/2]-1)
\\
-\Delta^{(\delta)}_{m-j \; j+1} & \quad (j=[m/2] , \cdots , m-1) ,
\end{array}
\right. 
\end{equation}
where $[m/2]$ stands for the integer part of $m/2$.
Putting (\ref{U-Delta}) and the symmetry,
\begin{equation}
\label{symmetry-UB}
\begin{array}{l}
U^{(\delta,m)}_{j} = U^{(\delta,m)}_{m-j}
\quad 
(j=0, 1, \dots , m-1)
\\
B^{(\delta,m)}_{n}=B^{(\delta,m)}_{m-n}
\quad (n=0, \cdots , m),
\end{array}
\end{equation}
together, we can rewrite (\ref{B-U-cont}) to as
\begin{equation}
\label{B-Delta}
\sum_{j=0}^{[m/2]-1}
\left( \sum_{n=j+1}^{m-(j+1)} B_{n}^{(\delta,m)} \right)
\Delta^{(\delta)}_{j+1 \, m-j} =0
\quad
(m=2, \cdots , \delta-1).
\end{equation}
From now on, we assume
\begin{equation}
\label{non-zero}
(v^{(\delta)}_0-v^{(\delta)}_2)^2 + (v^{(\delta)}_1)^2 \ne0
\end{equation}
on ${\mc M}(K^{(\delta)})$, which will not lose the generality
(see Appendix~\ref{proof-non-zero}). We show the following Lemma:
\begin{lemma}
\label{small-M}
Let ${\mc M}_{k}(K^{(\delta)})$ be the $2 \times k$ submatrix 
\begin{equation}
\label{sub-M}
{\mc M}_{k}(K^{(\delta)})
=
\left(
\begin{array}{cccc}
v^{(\delta)}_0 - v^{(\delta)}_2 & \cdots
& v^{(\delta)}_{k-1} - v^{(\delta)}_{k+1}
\\
v^{(\delta)}_1 & \cdots & v^{(\delta)}_{k}
\end{array}
\right) 
\quad
(k=1, \cdots , \delta-1)
\end{equation}
of ${\mc M}(K^{(\delta)})$ subject to (\ref{non-zero}).
If (\ref{B-Delta}) with $m=k+1$ hold true under
$\mbox{{\rm rank}} {\mc M}_{k}(K^{(\delta)}) = 1$,
then so does $\mbox{{\rm rank}} \, {\mc M}_{k+1}(K^{(\delta)}) = 1$.
\end{lemma}
%%%
\begin{proof}
We write down one of the assumption, (\ref{B-Delta}) with $m=k+1$,
more explicitly as
\begin{eqnarray}
\label{B-Delta-alt}
&&
\phantom{=}
\sum_{j=0}^{[\frac{k+1}{2}]-1}
\left( \sum_{n=j+1}^{(k+1)-j-1} B_{n}^{(\delta,k+1)} \right)
\Delta^{(\delta)}_{j+1 \, k+1-j}
\\
\nonumber
&&
=
\left( \sum_{n=1}^{(k+1)+1} B_{n}^{(\delta,k+1)} \right)
\Delta^{(\delta)}_{1 \, k+1}
+
\sum_{j=1}^{[\frac{k+1}{2}]-1}
\left( \sum_{n=j+1}^{(k+1)-j-1} B_{n}^{(\delta,k+1)} \right)
\Delta^{(\delta)}_{j+1 \, k+1-j}
=0.
\end{eqnarray}
Since the other assumption,
$\mbox{{\rm rank}} {\mc M}_{k}(K^{(\delta)}) = 1$, implies
\begin{equation}
\label{Delta-vanish}
\Delta_{j+1 \, k+1-j}=0 \quad (j=1, \cdots , [\textstyle{\frac{k+1}{2}}]-1),
\end{equation}
we can bring (\ref{B-Delta-alt}) into
\begin{equation}
\label{B-Delta-alt2}
\left( \sum_{n=1}^{(k+1)+1} B_{n}^{(\delta,k+1)} \right)
\Delta^{(\delta)}_{1 \, k+1}=0.
\end{equation}
so that we have $\Delta_{1 \, k+1} =0$.
The vanishment $\Delta_{1 \, k+1} =0$ is put together with
$\mbox{{\rm rank}} {\mc M}_{k}(K^{(\delta)}) = 1$ to show
$\mbox{{\rm rank}} \, {\mc M}_{k+1}(K^{(\delta)}) = 1$.
This completes the proof.
\end{proof}
%%%%%%%%%%%%%%
\indent
We are at the final stage to draw (\ref{deg-M}) from (\ref{nsc}):
Let us consider (\ref{B-Delta}) with $m=2$ for $K^{(\delta)}(q,p)$
subject to (\ref{non-zero}), which is written
explicitly as $\Delta^{(\delta)}_{1 \, 2} =0$. This shows
$\mbox{{\rm rank}} {\mc M}_{2}(K^{(\delta)})=1$ under (\ref{non-zero}).
We can thereby start applying Lemma~\ref{small-M} to (\ref{B-Delta})
recursively from $m=3$ to $m=\delta-1$,
and finally reach to
$\mbox{{\rm rank}} \, {\mc M}(K^{(\delta)})=\mbox{{\rm rank}}
\, {\mc M}_{(\delta-1)+1}(K^{(\delta)}) = 1$ under (\ref{non-zero}).
As for $K^{(\delta)}(q,p)$ not subject to (\ref{non-zero}), 
Appendix~\ref{proof-non-zero} shows that (\ref{deg-M}) is a necessary
condition of (\ref{nsc}).
Recalling Theorem~\ref{BDC-sep}, we have the following.
%%%%%%%%%
\begin{theorem}
\label{necessary}
If a $\delta$-PHO shares its BGNF up to degree-$(2\delta-2)$
with a $(2\delta-2)$-PHO up to degree-$(2\delta-2)$,
then the $\delta$-PHO is separable within a rotation
of Cartesian coordinates.
\end{theorem}
%%%%%%%%%
\par
Theorems~\ref{sufficient} and \ref{necessary} are put together to
make our conclusion: 
\par\smallskip
%%%%%%%%%
\noindent
{\sc Conclusion}.\quad
Our main theorem, Theorem~\ref{main}, holds true.
%%%%%%%%%%%%%%%%%%%%%%%%%%%%%%%%%%%%%%%%%%%%%%%%%%%%%
\appendix
%%%%%%%%%%%%%%%%
\section{Outline of the proof of Lemma~\ref{BG-norm-2(d-1)}}
\label{proof-lemma}
Since the BGNF of the $3$-PHO Hamiltonians has been given explicitly
\footnote{The expression agrees with (\ref{Gd<=2(d-1)})-(\ref{Gd-(2d-2)}).}
in \cite{U1}, we focus our attention only to the case of
$\delta \geq 4$ henceforth.
\par
(i)\quad
Equating the homogeneous parts of degree-$3$ on the both sides of
(\ref{def-BG-G}), we have
$
G^{(\delta)}_{3}(q,\eta)+(D^{(3)}_{q,\eta}W^{(\delta)}_{3})(q,\eta)
=
0 \, (=H^{(\delta)}_{3}(q,\eta))
$
.
On account of $\ker D^{(3)}_{q,\eta}=\{ 0 \}$
\footnote{There is no invariant homogeneous polynomials of odd-degree
under the $SO(2)$ action generated by $D_{q,\eta}$. {\it viz}
$\ker D^{(k)}_{q,\eta}=\{0\}$ for any odd $k$.}
, the equation above for the degree-$3$ part is solved as
$G^{(\delta)}_{3}(\xi,\eta)=0$ and $W^{(\delta)}_{3}(q,\eta)=0$.
Then by induction, we can show
\begin{equation}
\label{ell<d}
G^{(\delta)}_{\ell}(\xi,\eta)=0 \quad W^{(\delta)}_{\ell}(q,\eta)=0
\quad (\ell = 3, \cdots , \delta-1).
\end{equation}
Under (\ref{ell<d}), the degree-$\delta$ part of (\ref{def-BG-G})
takes the form
$
G^{(\delta)}_{\delta}(q,\eta)
+
(D^{(\delta)}_{q,\eta}W^{(\delta)}_{\delta})(q,\eta)
=
V^{(\delta)}(q) \, (=H^{(\delta)}_{\delta}(q,p)),
$
so that we have
\begin{equation}
\label{ell=d}
G^{(\delta)}_{\delta}(\xi,\eta)= V^{(\delta)}{}^{\ker}(\xi,\eta)
\quad
W^{(\delta)}_{\delta}(q,\eta)
=
({\tl D}^{(\delta)}_{q,\eta}{}^{-1}V^{(\delta)}{}^{\supimage})(q,\eta)
\end{equation}
where ${\tl D}^{(\delta)}_{q,\eta}$ denotes the restrict of
$D^{(\delta)}_{q,\eta}$ on its image. This shows (\ref{Gdd}).
\par
(ii) \quad Under (\ref{ell<d}) and (\ref{ell=d}), we can show
\begin{equation}
\label{d<ell<2d-2}
G^{(\delta)}_{\ell}(\xi,\eta)=0 \quad W^{(\delta)}_{\ell}(q,\eta)=0
\quad (\ell = \delta+1, \cdots , 2\delta-3)
\end{equation}
by induction. On substituting (\ref{ell<d})-(\ref{d<ell<2d-2}) into
(\ref{def-BG-G}), the degree-$(2\delta-2)$ part of (\ref{def-BG-G})
is calculated to be
\begin{equation}
\label{ell=2d-2}
G^{(\delta)}_{2\delta-2}(q,\eta)
+
(D^{(\delta)}_{q,\eta}W^{(\delta)}_{2\delta-2})(q,\eta)
=
\frac{1}{2} \sum_{j=1}^{2} \left\{
\left( \frac{\partial W^{(\delta)}_{\delta}}{\partial q_{j}} \right)^2
- 
\left( \frac{\partial W^{(\delta)}_{\delta}}{\partial \eta_{j}} \right)^2
\right\} .
\end{equation}
The final expression (\ref{Gd-(2d-2)}) with (\ref{coeff-Gd})
is obtained by writing down explicitly the kernel component of
the rhs of (\ref{ell=2d-2}),
which requires another very simple but long calculation.
%%%%%%%%%%%%%%%%%%%%%%%%%
\section{Proof of (\ref{vanish-sum-B})}
\label{proof-vanish}
From (\ref{B}), the rhs of (\ref{vanish-sum-B}) is put in a form
\begin{eqnarray}
\label{sort-sum-B}
&&
\phantom{=}
\displaystyle{
\left( (\delta-1)^2 \textstyle{{\delta-2} \choose {m-1}} \right)^{-1} 
\sum_{n=0}^{m} B_{n}^{(\delta,m)}
}
\\
\nonumber
&&
=
\displaystyle{
\sum_{k=0}^{\delta-2}
\frac{ {{\delta-2} \choose k} }{ \{2(k+1)-\delta\} (2k-\delta) }
\left( \sum_{n=N_1^{\flat}}^{N_1^{\sharp}}
\textstyle{ {{m-1} \choose n} {{\delta-m} \choose {n+k+1-m}} }
\right)
}
\\
\nonumber
&&
\phantom{==}
+
\displaystyle{
\sum_{k=0}^{\delta-2}
\frac{ {{\delta-2} \choose k} }{ \{2(k+2)-\delta \} \{2(k+1)-\delta \} }
\left( \sum_{n=N_{2}^{\flat}}^{N_{2}^{\sharp}}
\textstyle{ {{m-1} \choose {n-1}} {{\delta-m} \choose {n+k+1-m}} }
\right)
}
\\
\nonumber
&&
\phantom{====}
+
\displaystyle{
\sum_{k=0}^{\delta-1}
\frac{ {{\delta-1} \choose k} }{ \{2(k+1)-\delta \} (2k-\delta)}
\left(\sum_{n=N_3^{\flat}}^{N_3^{\sharp}}
\textstyle{ {{m} \choose {n}} {{\delta-m-1} \choose {n+k-m}} }
\right)
},
\end{eqnarray}
where
\begin{equation}
\label{range-N}
\begin{array}{ll}
N_1^{\flat}=\max (0, m-1-k) \quad & N_1^{\sharp}=\min (m-1, \delta-1-k)
\\
N_2^{\flat}=\max (1, m-1-k) \quad & N_2^{\sharp}=\min (m, \delta-1-k)
\\
N_3^{\flat}=\max (0, m-k)   \quad  & N_3^{\sharp}=\min (m, \delta-1-k).
\end{array}
\end{equation}
For the sums with the summation index $n$ on the rhs of (\ref{sort-sum-B}),
we have
\begin{equation}
\label{binom}
\begin{array}{l}
\displaystyle{\sum_{n=N_1^{\flat}}^{N_1^{\sharp}}}
\textstyle{ {{m-1} \choose n} {{\delta-m} \choose {n+k+1-m}} }
=
\displaystyle{\sum_{n=N_3^{\flat}}^{N_3^{\sharp}}}
\textstyle{ {{m} \choose {n}} {{\delta-m-1} \choose {n+k-m}} }
=
\textstyle{{{\delta-1} \choose k}}
\\
\displaystyle{
\sum_{n=N_{2}^{\flat}}^{N_{2}^{\sharp}}}
\textstyle{ {{m-1} \choose {n-1}} {{\delta-m} \choose {n+k+1-m}} }
=
\textstyle{{{\delta-1} \choose {k+1}}}.
\end{array}
\end{equation}
Hence, (\ref{sort-sum-B})-(\ref{binom}) are put together to show
\begin{eqnarray}
\label{cal-sum-B}
&&
\frac{ \displaystyle{\sum_{n=0}^{m} B_{n}^{(\delta,m)}} }
     { (\delta-1)^2 \textstyle{{\delta-2} \choose {m-1}} }
=
\left[
\displaystyle{
\sum_{k=0}^{\delta-2}
}
\frac{ {{\delta-1} \choose k} {{\delta-2} \choose k} }
     { \{2(k+1)-\delta\} (2k-\delta) }
\right.
\\
\nonumber
&&
\phantom{\displaystyle{
\sum_{n=0}^{m} B_{n}^{(\delta,m)}======}}
+ 
\displaystyle{
\sum_{k=0}^{\delta-2}
}
\frac{ {{\delta-1} \choose {k+1}} {{\delta-2} \choose k} }
     { \{2(k+2)-\delta \} \{2(k+1)-\delta \} }
\\
\nonumber
&&
\phantom{\displaystyle{
\sum_{n=0}^{m} B_{n}^{(\delta,m)}==========}}
+
\left.
\displaystyle{
\sum_{k=0}^{\delta-1}
}
\frac{ {{\delta-1} \choose k}{{\delta-1} \choose k} }
     { \{2(k+1)-\delta \} (2k-\delta)}
\right]
\\
\nonumber
&&
\phantom{
\frac{ \displaystyle{\sum_{n=0}^{m} B_{n}^{(\delta,m)}} }
     { (\delta-1)^2 \textstyle{{\delta-2} \choose {m-1}} }
}
=
\displaystyle{\sum_{k=1}^{\delta-2}}
\frac{ {{\delta -1} \choose k} 
       \{ {{\delta-2} \choose k} + {{\delta-2} \choose {k-1}}
              - {{\delta-1} \choose k} \} }
     { \{2(k+1)-\delta \}(2k-\delta) } =0.
\end{eqnarray}
%%%%%%%%%%%%%%%%%%%%%%%
\section{$SO(2)$ action to ${\mc M}(K^{(\delta)})$}
\label{proof-non-zero}
To make the following discussion simple, we assume that $\delta$ is odd.
Let $K^{\prime}{}^{(\delta)}(q,p)$ be the $\delta$-PHO Hamiltonian given by
\begin{equation}
\label{Kd'}
K^{\prime}{}^{(\delta)}(q,p)
=
K^{(\delta)}(\sigma(\psi)^{-1}q,\sigma(\psi)^{-1}p),
\end{equation}
where $K^{(\delta)}(q,p)$ is defined by (\ref{Kd}), and
$\sigma(\psi)$ by (\ref{rot-mat}).
It is then shown by a straightforward calculation that they are related
as
\begin{equation}
\label{SO2-Kd}
{\mc M}(K^{\prime}{}^{(\delta)})
=
\sigma(2\psi) {\mc M}(K^{(\delta)}) R^{(\delta-2)}(\psi),
\end{equation}
where $R^{(\delta-2)}(\psi)$ is the standard representation of $SO(2)$
on the real vector space of homogeneous polynomials of degree-$(\delta-2)$.
\par
Assume that $K^{(\delta)}(q,p)$ is not subject to (\ref{non-zero}).
What we have to show is the existence of $\sigma (\psi)$
which brings $K^{(\delta)}(q,p)$ to $K^{\prime}{}^{(\delta)}(q,p)$
with ${\mc M}(K^{\prime}{}^{(\delta)})$ having non-vanishing first column.
Since the non-existence of such $\sigma(\psi)$ is equivalent
to that the vector subspace, ${\mc N}= \mbox{{\rm span}}
\{ q_1^{h}q_2^{\delta-2-h} \}_{h=1, \cdots, \delta-2}$, is an invariant
subspace of the $SO(2)$ action given by $R^{(\delta-2)}(\psi)$.
However, this is not true: It is easily seen that any
invariant subspace is given by a direct sum of the $2$-dimentional
subspaces each of which is spanned by
$\Re ((q_1+iq_2)^{h}(q_1-iq_2)^{\delta-2-h})$
and $\Im ((q_1+iq_2)^{h}(q_1-iq_2)^{\delta-2-h})$
($h=0, \cdots , (\delta-1)/2$). Hence ${\mc N}$ is not $SO(2)$-invariant.
\par
We are now at the final stage to explain that the assumption (\ref{non-zero})
does not lose generality of our proof of the necessity.
Let consider the $\delta$-PHO Hamiltonian $K^{(\delta)}(q,p)$
which is not subject to (\ref{non-zero}) and shares its BGNF with
a $(2\delta-2)$-PHO $K^{(2\delta-2)}(q,p)$.
As shown above, we can bring $K^{(\delta)}(q,p)$
to $K^{\prime}{}^{(\delta)}(q,p)$ through (\ref{Kd'}) that satisfies
(\ref{non-zero}) by adopting a suitable rotation with $\sigma(\psi)$.
According to the commutativity of the BG normalization and
the rotations shown in Section~3, $K^{\prime}{}^{(\delta)}(q,p)$
shares its BGNF with a $(2\delta-2)$-PHO $K^{\prime}{}^{(2\delta-2)}(q,p)$
other than $K^{(2\delta-2)}(q,p)$. Hence, the discussion
in subsection.~4.2 is applied to $K^{\prime}{}^{(\delta)}(q,p)$
to show the necessity of $\mbox{rank} {\mc M}(K^{\prime}{}^{(\delta)})=1$.
Accordingly, the equation,
\begin{equation}
\label{rank-Kd-Kd'}
\mbox{rank} {\mc M}(K^{\prime}{}^{(\delta)})
= \mbox{rank} {\mc M}(K^{(\delta)}),
\end{equation} 
following from (\ref{SO2-Kd}) leads us to (\ref{deg-M}).
%%%%%%%%%%%%%%%%%%%%%%%%%%%%%%%%%%%%%%%%%%%%%%%%%%%%%%%%
\bibliographystyle{amsalpha}

\begin{thebibliography}{UCRV}
\bibitem [A]{A} V. I. Arnold, \textit{Mathematical methods of Classical
mechanics}, Springer-Verlag, New York, NY, 1980.
%
\bibitem [B]{Ber} J. Bertrand, J.~de~ Math. (i), {\bf 17} (1852),
121.
%
\bibitem [D]{D} G. Darboux, Arch. Neerlandaises ii, {\bf 6} (1901),
371.
%
\bibitem [G]{G} H. Goldstein, {Classical mechanics}, Addison-Wesley,
Reading, MA, 1950.
%
\bibitem [GPS]{GPS} C. Grosche, G. S. Pogosyan and A. N. Sissakian,
Frotschr. Phys., {\bf 43} (1995), 453.
%
\bibitem [H]{H} J. Hietarinta, Phys. Rep., {\bf 147} (1987), 87.
%
\bibitem [M]{M} J. K. Moser, \textit{Lectures on Hamiltonian systems},
Memoires of AMS vol.81, Amer. Math. Soc., Providence, RI, 1968.
%
\bibitem [MW]{MW} I. Marshall and S. Wojciechowski,
J. Math. Phys., {\bf 29} (1988), 1338.
%%
\bibitem [UCRV]{UCRV} Y. Uwano, N. Chekanov, V. Rostovtsev and S. Vinitsky,
\textit{Computer algebra in scientific computing} (ed. V. G. Ghanza et a),
Springer-Verlag, Berlin, 1999, 441.
%
\bibitem [U1]{U1} Y. Uwano, J. Phys. {\bf A}, {\bf 33} (2000), 6635.
%
\bibitem [U2]{U2} Y. Uwano, \textit{Separability of the perturbed
harmonic oscillators with homogeneous polynomial potentials
from the Birkhoff-Gustavson normalization viewpoint I, II},
in preparation.
\bibitem [W]{W} E. T. Whittaker, \textit{A treatise on the analytical
dynamics of particles and rigid bodies}, Cambridge U. P., Cambridge,
1937.
\end{thebibliography}

\end{document}